\documentclass[a4paper]{jpconf}
\usepackage{graphicx}
\usepackage{amsmath,amsthm,amsfonts}
\usepackage{amssymb,epsf}
\usepackage{latexsym}

\def\us{\char`\_}

\newcommand{\insertplot}[5]{\begin{figure}
 \hfill\hbox to 0.05in{\vbox to #5in{\vfill
 \inputplot{#1}{#4}{#5}}\hfill}
 \hfill\vspace{-.1in}
 \caption{#2}\label{#3}
 \end{figure}}
 \newcommand{\inputplot}[3]{
 \special{ps: plotfile #1}
}
\newcounter{fig}

\newcommand{\vphi}{\varphi}
\newcommand{\vepsilon}{\varepsilon}

\begin{document}
\title{Charged Rotating Black Holes in Higher Dimensions}

\author{Masoud Allahverdizadeh${}^{1}$, Jutta Kunz${}^{1}$,  Francisco Navarro-L\'erida${}^{2}$}

\address{
${}^{1}$Institut f\"ur Physik, Universit\"at Oldenburg, Postfach 2503, 
D-26111 Oldenburg, Germany. \\
${}^{2}$Departamento de F\'{\i}sica At\'omica, Molecular y Nuclear, Ciencias F\'{\i}sicas,\\
Universidad Complutense de Madrid, E-28040 Madrid, Spain.
}

\ead{masoud\us alahverdi@yahoo.com, kunz@theorie.physik.uni-oldenburg.de, fnavarro@fis.ucm.es}

\begin{abstract}
In recent years higher-dimensional black holes have attracted 
much interest because of
various developments in gravity and high energy physics. 
But whereas higher-dimensional charged static (Tangherlini) 
and uncharged rotating (Myers-Perry) black holes were found long ago, 
black hole solutions of Einstein-Maxwell theory,
are not yet known in closed form in more than 4 dimensions,
when both electric charge and rotation are present. 
Here we therefore study these solutions
and those of Einstein-Maxwell-dilaton theory,
by using numerical and perturbative methods, 
and by exploiting the existence of spacetime symmetries.
The properties of these black holes reveal new interesting
features, not seen in $D=4$. 
For instance, unlike the $D=4$ Kerr-Newman solution,
they possess a non-constant gyromagnetic factor.
\end{abstract}

\section{Introduction}

In Einstein-Maxwell (EM) theory in $3+1$ dimensions
the unique family of stationary asymptotically flat black holes with
non-degenerate horizon 
comprises the rotating Kerr-Newman (KN) and Kerr black holes
and the static Reissner-Nordstr\"om (RN) and Schwarzschild black holes.
EM black holes are uniquely determined
by their mass, their electric and magnetic charge,
and their angular momentum
(see \cite{unique} and references therein).

The generalization of black hole solutions to higher
dimensions was pioneered by Tangherlini
\cite{tangher} for static vacuum black holes,
and by Myers and Perry (MP) \cite{MP} for stationary vacuum black holes.
Whereas Myers and Perry \cite{MP} also obtained
charged static black holes in higher dimensional EM theory,
higher dimensional charged rotating black holes have not yet
been obtained in closed form in pure EM theory, 
although such black holes are known
for some low energy effective actions related to string theory (including
additional fields, though) 
\cite{string1a,string1b,string2a,string2b}.

Higher dimensional black holes received much interest in recent years also
with the advent of brane-world theories, raising the possibility
of direct observation in future high energy colliders
\cite{exp}, which makes the theoretical study of these higher dimensional
objects particularly appealing.

However, finding black holes solutions in higher dimensions is a difficult task due to
the size and complexity of the equations. Moreover, contrary to what happens
in $D=4$ the topology of the horizon is not unique in higher dimensions (for
instance, ringlike configurations are allowed \cite{Emparan}), which makes the
problem even harder. For these reasons the number of solutions known analytically
in closed form is very limited.
This forces us to use alternative techniques to study
these black holes. Here we will briefly analyze 
such black holes by two methods: the numerical
method and
the perturbative method.

\section{Einstein-Maxwell-Dilaton theory}
We will concentrate on Einstein-Maxwell-Dilaton (EMD) theory in $D$
dimensions, as an example how these methods may be applied. The classical EMD action reads
\begin{eqnarray}
S &=&\int d^{D}x\sqrt{-g}\left(
R\text{
}-\frac{1}{2}\partial_{\mu}\Phi \partial^{\mu}\Phi-\frac{1}{4}e^{-2h \Phi}F_{\mu \nu }F^{\mu \nu }\right),  \label{act1}
\end{eqnarray}
where ${R}$ is the scalar curvature, $\Phi$ is the dilaton field,
$F_{\mu \nu }=\partial _{\mu }A_{\nu }-\partial _{\nu }A_{\mu }$
is the electromagnetic field tensor, and $A_{\mu }$ is the
electromagnetic potential. $h$ is the dilaton coupling constant.

The equations of motion can be obtained by varying the action 
with respect to the metric $g_{\mu \nu}$, the dilaton field $\Phi $  
and the gauge potential $A_{\mu }$, yielding
\begin{equation}
G_{\mu\nu}=\frac{1}{2}\left[\partial_{\mu}\Phi \partial_{\nu}\Phi
-\frac{1}{2}g_{\mu \nu}\partial_{\rho}\Phi \partial^{\rho}\Phi
+e^{-2h \Phi}\left(F_{\mu\rho} {F_\nu}^\rho 
 - \frac{1}{4} g_{\mu \nu} F_{\rho \sigma} F^{\rho \sigma}\right)\right]
\ , \label{FE1}
\end{equation}
\begin{equation}
\nabla ^{2}\Phi   =-\frac{h}{2}e^{-2h \Phi}F_{\mu\nu} F^{\mu \nu} \ , \ \ \  \nabla_\mu \left(e^{-2h \Phi}F^{\mu \nu}\right)  = 0 \ . \
\label{FE3}
\end{equation}
For $h=0$ the theory reduces to pure EM theory.

These equations admit an analytical solution when the dilaton
coupling constant $h$ takes the Kaluza-Klein (KK) value $h_{KK}$ \cite{Llatas,Kunz1}
\begin{equation}
h_{\rm KK}=\frac{D-1}{\sqrt{2(D-1)(D-2)}} \ . \label{h_def}
\end{equation}

In order to generate charged rotating solutions for generic values of $h$ we will assume that
the topology of the horizon is spherical. 
We will further assume that the black holes are stationary and
axisymmetric.
Stationary axisymmetric black holes in $D$ dimensions possess $N = [(D-1)/2]$
independent angular momenta $J_{i}$ associated with
$N$ orthogonal planes of rotation \cite{MP}.
($[(D-1)/2]$ denotes the integer part of $(D-1)/2$,
corresponding to the rank of the rotation group $SO(D-1)$.)
These black hole solutions then fall into two classes:
 even-$D$ and odd-$D$-solutions.

When all $N$ angular momenta have equal-magnitude,
the symmetry of these black holes is strongly enhanced.
In fact, in odd dimensions, the symmetry then increases from
$R \times U(1)^N$ to $R \times U(N)$,
thus changing the problem from cohomogeneity-$N$ to cohomogeneity-1.
As a consequence the original system of partial differential equations
in $N$ variables then reduces to a system of ordinary differential
equations (ODE's).

\section{Numerical and perturbative methods}
In what follows we will focus on the case where $D$ is odd, all angular
momenta have the same magnitude, and the topology of the horizon is that of a
sphere. That case is tractable numerically and perturbatively. Under those
assumptions the metric, the gauge potential and the dilaton field  can be
parametrized with their angular dependence explicitly known
\begin{eqnarray}
&&ds^2 = \frac{m}{f} \left[ dr^2 + r^2 \sum_{i=1}^{N-1}
  \left(\prod_{j=0}^{i-1} \cos^2\theta_j \right) d\theta_i^2\right] +\frac{n}{f} r^2 \sum_{k=1}^N \left( \prod_{l=0}^{k-1} \cos^2 \theta_l
  \right) \sin^2\theta_k \left(\vepsilon_k d\vphi_k - \frac{\omega}{r}
  dt\right)^2 \nonumber \\
&&+\frac{m-n}{f} r^2 \left\{ \sum_{k=1}^N \left( \prod_{l=0}^{k-1} \cos^2
  \theta_l \right) \sin^2\theta_k  d\vphi_k^2 \right. -\left. \left[\sum_{k=1}^N \left( \prod_{l=0}^{k-1} \cos^2
  \theta_l \right) \sin^2\theta_k \vepsilon_k d\vphi_k\right]^2 \right\} -f dt^2  \ , \label{metric}
\end{eqnarray}
\begin{equation}
A_\mu dx^\mu =  a_0 dt + a_\vphi \sum_{k=1}^N \left(\prod_{l=0}^{k-1}
  \cos^2\theta_l\right) \sin^2\theta_k \vepsilon_k d\vphi_k \ , \ \ \ \Phi=\Phi(r) \ , \label{gauge_Phi}
\end{equation}
where $\theta_0 \equiv 0$, $\theta_i \in [0,\pi/2]$ 
for $i=1,\dots , N-1$, 
$\theta_N \equiv \pi/2$, $\vphi_k \in [0,2\pi]$ for $k=1,\dots , N$,
and $\vepsilon_k = \pm 1$ denotes the sense of rotation
in the $k$-th orthogonal plane of rotation. The spacetime possesses $N+1$ commuting Killing vectors,
$\xi \equiv \partial_t$, 
and $\eta_{(k)} \equiv \partial_{\vphi_k}$, for $k=1, \dots , N$.
This parametrization employs
isotropic coordinates and is appropriate for numerical purposes
\cite{Kunz2,Kunz3}. However, for perturbative purposes Boyer-Lindquist
coordinates simplify the algebra \cite{Masoud1,Masoud2}. 

In order that the solutions correspond to black holes they must possess an
event horizon located at $r=r_{\rm H}$, characterized by the condition $f(r_{\rm
  H})=0$ \cite{kkrot}. Such an event horizon is in fact a Killing horizon
since the  Killing vector
\begin{equation}
\chi = \xi + \Omega \sum_{k=1}^N \vepsilon_k \eta_{(k)} \ ,
\label{chi} \end{equation}
 is null at and orthorgonal to the horizon, $\Omega$ representing the constant
 horizon
 angular velocity. 

The system of ODE's resulting when
Eqs.~(\ref{metric}-\ref{gauge_Phi}) are substituted in Eqs.~(\ref{FE1}-\ref{FE3})
must be supplemented with the appropriate set of boundary conditions. This is
obtained by imposing that the horizon is regular and the spacetime is
asymptotically flat \cite{Kunz1,Kunz2,Kunz3}. Then the system of ODE's
 is solved by employing a numerical solver (e.g., COLSYS), 
providing accurate results. The numerical parameters are $\{ h,
r_{\rm H}, \Omega , Q\}$, with $Q$ being the electric charge. Varying these
parameters we obtain families of solutions, with their corresponding values of
the physical properties like the mass $M$, the angular momentum $J$, etc. An
important information the numerical method provides is the domain of existence
of black holes in higher dimensions. That domain consists of the region in the
parameter space bounded by the extremal solutions. In Figure 1 (left) we exhibit the
domain of existence of EMD black holes in terms of the scaled electric charge $Q/M$ and
angular momentum $J/M^{(D-2)/(D-3)}$. Clearly the domain depends both on the
dilaton coupling constant $h$ and the dimension $D$.

\begin{figure}[h!]
\parbox{\textwidth}
{\centerline{
\mbox{
\epsfysize=5.0cm
\includegraphics[width=60mm,angle=270,keepaspectratio]{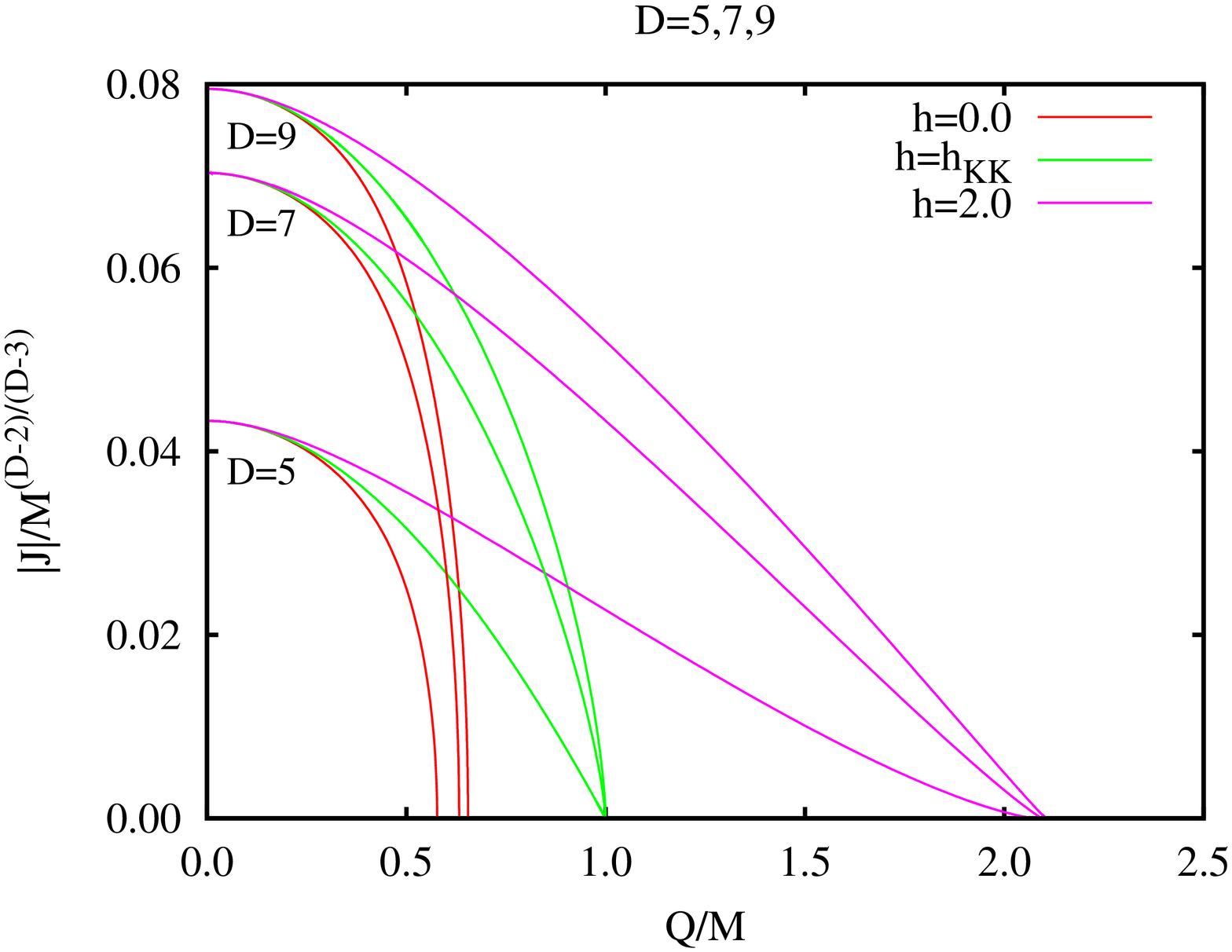}
\includegraphics[width=60mm,angle=270,keepaspectratio]{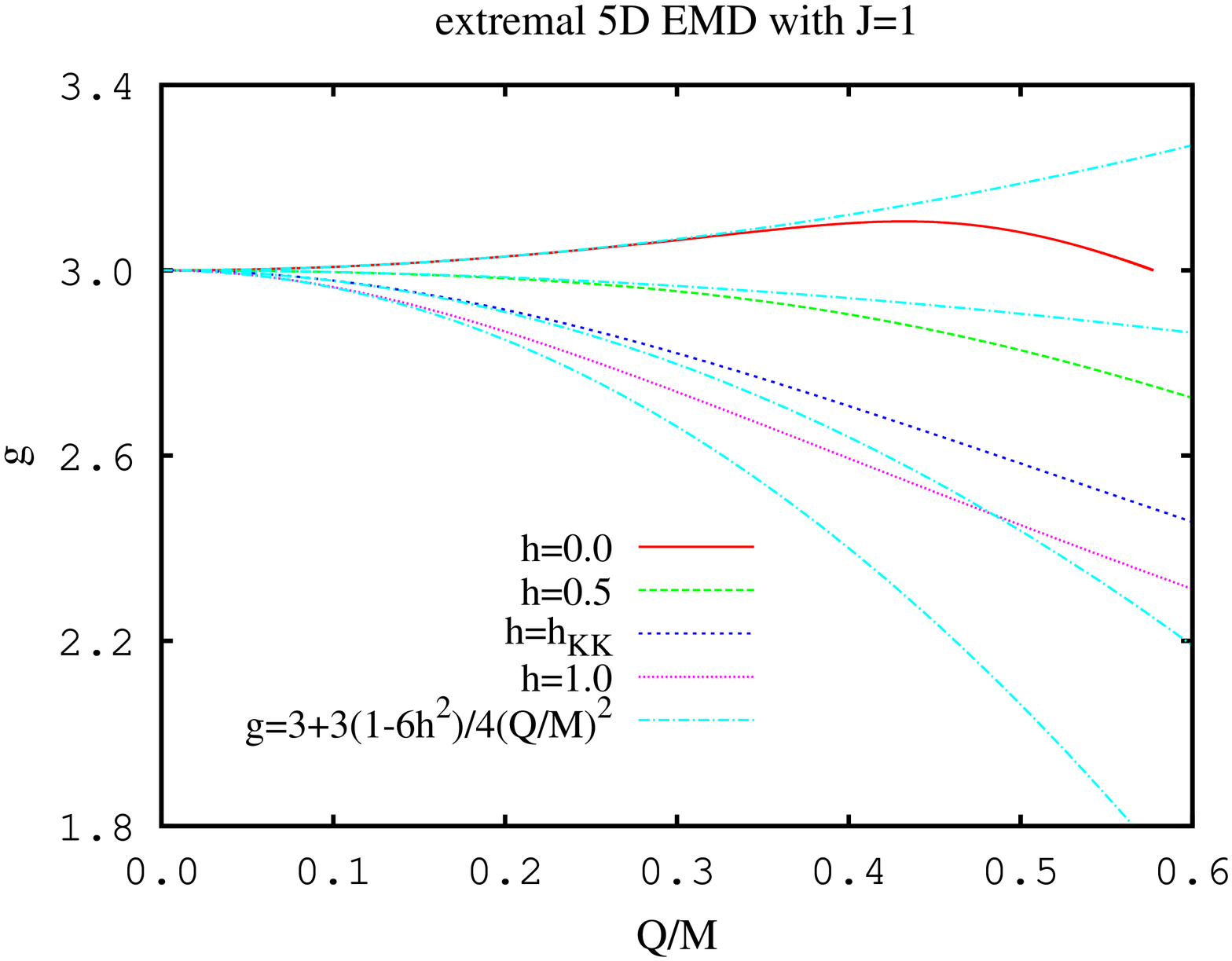}
}}}
\caption{
Left: Domain of existence of EMD black holes 
with equal-magnitude angular momenta
in $D=5$, 7 and 9 dimensions.
Right: Comparison of the gyromagnetic factor $g$ of $D=5$ extremal EMD black
holes obtained with numerical and perturbative (cyan)  methods.
}
\end{figure}

The accuracy of the numerical method may be tested by means of the analytical
KK solutions, as well as by exact relations (like the Smarr mass formula) these
black holes satisfy \cite{Masoud2}.

As a complementary method one may employ perturbations to generate black hole
solutions in higher dimensions \cite{Masoud1, Masoud2, Navarro}. The advantage of this method is that it
produces analytical formulae, which the numerical method does not. However,
the range of validity of these formulae is limited by the accuracy of the order
of the perturbations.

A good perturbative parameter to produce charged rotating black hole solutions
in higher dimensions is the electric charge. One starts from the MP solution
\cite{MP} and expands the metric, the gauge potential and the dilaton in series
expansions in the electric charge around the MP solution (see \cite{Masoud2} for details). Those
series are substituted in Eqs.~(\ref{FE1}-\ref{FE3}) and solved order
by order in the electric charge. For lower orders the equations are easy to
solve but as the order of the perturbations increases the equations become
more and more difficult to solve. 

An important point when applying the perturbative method is the way the
integration constants are fixed. One has to impose regularity of the solutions
at the horizon together with the condition of asymptotic flatness. The
Smarr mass formula is 
a useful means to check the accuracy of the solutions.

When the perturbative method is applied to extremal solutions the formulae
simplify very much. For instance, one obtains the following formula of the
gyromagnetic factor $g$ to second order
\begin{equation}
\frac{g}{D-2} = 1 + \frac{1}{4} \left[ 1 - \frac{2(D-1)(D-2)}{(D-3)^2} h^2
\right] \left(\frac{Q}{M}\right) ^2 + o\left(\left(\frac{Q}{M}\right)^2\right)
\ , \label{gfactor}
\end{equation}
showing that the gyromagnetic factor is not constant for pure EM solutions ($h=0$).

\section{Comparison of the two methods} 

Although both methods have a different range of applicability there is a region of the
parameter space where both are valid, namely, when the electric charge is small.
In that region one can check the consistency of
the methods and their level of agreement \cite{Masoud2}. This is presented in
Figure 1 (right) where the perturbative formula Eq.~(\ref{gfactor}) for extremal
solutions is compared to the corresponding non-perturbative numerical results
for $D=5$. We observe that for $|Q/M| \le 0.2$ both approches are in good
agreement. In fact, the perturbative method may be modified to analyze also
slowly rotating solutions, using as the unperturbated initial solution the
Tangherlini solution and using the angular momentum as the perturbative
parameter. 

To summarize, these two methods are complementary and provide us with a
powerful tool to analyze black holes in higher dimensions when analytical
solutions are not available.

\end{document}